# Higgs Factory and 100 TeV Hadron Collider: Opportunity for a New World Laboratory within a Decade

Peter McIntyre[1], Saeed Assadi[1], Chase Collins[1], James Gerity[1], Joshua Kellams[1], Thomas Mann[1], Christopher Mathewson[1], Nathaniel Pogue[1], Akhdiyor Sattarov[1], and Richard York[2]

Suggestions have been made for a 80-100 km circumference Future Circular Collider (FCC) that could ultimately contain a circular $e^+e^-$ ring collider operating as a Higgs Factory as well as a 100 TeV hadron collider [1]. Those suggestions have motivated us to propose a World Laboratory to achieve those goals and more, with minimum cost and risk, building upon three ingredients:

- a site with existing tunnels and favorable geotechnology for minimum-cost tunneling;
- use of low risk/low cost magnets that is possible only with a large-circumference site, and
- the possibility that Texas might offer to provide the tunnels as cost-sharing.

## Higgs Factory

**Higgs Factory in the SSC Tunnel.** The 87 km circumference SSC tunnel was ~45% complete at the termination of the project, and the linac and low-energy booster tunnels were completed. We suggest the possibility of a Cooperative Agreement in which the State of Texas would provide the site and complete the tunnels necessary for a Higgs Factory and its injectors, and DOE would fund the construction of the technical systems for the project.

**Dual-use Injector: Top-up injection to the Higgs Factory, X-ray FEL for Structural Systems Biology.** The Higgs Factory will require frequent top-up injection, which will require a ~9 GeV linac, a positron source and damping ring, and sequential acceleration to full beam energy. The 9 GeV initial acceleration for electrons and positrons can be most cost-effectively provided using a recirculating linac with 4 passes. York has suggested that the same approach could provide a cost-effective basis for an X-ray FEL capable of the short-pulse high-energy X-ray bunches needed for structural systems biology [2]. We suggest that it might be possible to secure funding of the XFEL as a cooperation among the State of Texas and life sciences foundations and agencies. The dual-use application offers the potential to launch world-class science at the facility early in the project even while the Higgs Factory is being built.

## Hadron Collider

**Minimum-cost route to 100 TeV: Large-circumference, modest magnetic field.** The SSC tunnel set multiple world records for tunnel-boring advance rate and had the lowest cost/meter of any large-bore tunnel. That is a consequence of its location in two rock strata, the Austin Chalk and the Taylor Marl, soft consolidated Cretaceous rock that is an ideal medium for large-bore tunneling. We have established that a second 270 km circumference tunnel could be bored in the same favorable rock strata to house the 100 TeV hadron collider. That large circumference would correspond to 5 Tesla superconducting magnets for a 100 TeV hadron collider. Below we present an example of a dual-bore magnet suitable for that collider that would minimize the cost and eliminate the technology risk for the collider rings. The 270 km tunnel could be aligned tangent to the SSC tunnel so that it could house the injector accelerator.

[1]Dept. of Physics, Texas A&M University   [2]Dept. of Physics, Michigan State University

**Future upgrade to 300 TeV.** Choosing a modest magnetic field and large circumference for a hadron collider not only minimizes its cost and risk; it also provides a future opportunity to upgrade the collider if/when the day comes that high-field magnet technology is affordable. The 15-20 T magnets that have been suggested for a 80-100 km circumference collider could be then installed in the 270 km tunnel to achieve 300 TeV collision energy.

The Higgs Factory project requires only the completion of the SSC tunnel for its home; the 100 TeV Hadron Collider requires the SSC tunnel for its injector and a new 270 km tunnel. In what follows we will show that each project is minimum-cost and minimum-risk in the proposed approach, whether they are undertaken together or independently.

## Leveraging State-of-Art RF Technology for the Higgs Collider

For the superconducting RF (SRF) systems we choose an SRF frequency of 1500 MHz for the SRF injector linac, and 750 MHz for the SRF acceleration that is used to replace energy loss from synchrotron radiation in the collider. After the recirculating linac the beam is distributed using an RF dipole [3] on a bunch-by-bunch basis to produce and inject positrons and electrons to the Higgs factory. This occurs only a small fraction of the time, while >90% of the time the linac drives the FEL. The choice of 1500 MHz enables us to utilize the new JLab C-100 cryomodules, built for their 12 GeV upgrade. It is the fruit of a decade-long development and enables us reduce development costs and to be ready to begin construction of the cryomodules for the XFEL injector complex just as soon as the frequency choice is validated in the accelerator design. The recirculating linac will make use of the existing SSC linac tunnel [4] and beam transfers will utilize existing beam lines where appropriate, reducing the conventional facility cost for the injector construction.

## Minimizing the cost of tunnels for the Higgs Factory and the Hadron Collider

Both the Higgs Factory and the hadron collider tunnel costs will be substantially reduced because of the favorable geotechnology in the region of the SSC tunnel. In addition, we would envisage proposing to the State of Texas that it complete the civil construction for the injector complex, the 87 km SSC tunnel for the Higgs Factory, and a 270 km tunnel for the 100 TeV hadron collider as cost-sharing support for the project.

The dominant capital costs for the 100 TeV hadron collider are the tunnel, the double-ring of superconducting magnets, and the liquid helium refrigerators. The dominant operating cost is the liquid He refrigeration. The choice of tunnel circumference optimizes oppositely for these several cost drivers. Lower-field dipole technology has substantially lower cost/TeV than higher-field dipole technology, and total synchrotron radiation power scales inversely with radius. An overall optimization thus requires that one balance the capital cost of the technical systems and lifetime operating cost against the tunnel cost.

For a site adjacent to CERN the proposed alignment encircles the Rhone valley and the Saleve because the mountains limit the circumference to a maximum of ~100 km. The tunnel for that alignment must pass through many rock strata and under the river and the lake, and would likely present a high construction cost/m. Two benchmarks for alpine tunnels are the 3 m diameter LEP tunnel cost ~11,000CHF/m in 1981 [5] (equivalent to $17,300/m in today's money), and the recent Gothard road tunnel, containing two 8 m diameter tunnels each 57 km length with a cost of 11.8 billion CHF [6], which would scale by area to a cost of $23,000/m for a 4 m diame-

ter tunnel. By contrast the 4.2 m diameter SSC tunnel cost ~$3,000/m in 1992, which would escalate to $4,900/m today. One can conclude that the cost ratio of mixed-rock alpine tunneling to homogeneous soft-rock tunneling is a factor ~4.

We report an alternative siting that would use to advantage the existing tunnel built for the SSC (Figure 1). The SSC tunnel boring advance rate retains the world records for best day, week, and month [7], and had a low construction cost: the 4.2 m-bore tunnels had a consistent cost of $2400/m for segments entirely within the Austin Chalk and Taylor Marl, $4000/m for segments that spanned between Austin Chalk and Eagle Ford Shale [8]. Figure 1b shows a location where the tunnel passed through an inactive fault. Even at the fault location the record tunnel advance rates were sustained. At the time the SSC project was terminated ~39 km of tunnel had been bored, of which ~30% had been lined by the termination of the project. The un-lined tunnel sections should be intact today as when they were completed. There are many un-lined large-bore tunnels in the Austin Chalk that have been in service for many years as water diversion channels. Figure 2 shows a geologic map and stratigraphy for the SSC tunnel and for a possible position for a future 270 km-circumference tunnel that would house the 100 TeV hadron collider. The large tunnel is planar, stays entirely within the favorable rock strata, and should require no shafts deeper than ~200 m.

The SSC tunnel would house the Higgs Factory implementation of the FCC design [9], with collision energy of 240 GeV. It would also house a 15 TeV proton injector for the hadron collider. Such a high-energy injector would reduce the dynamic range over which the collider magnets must operate to $B_{max}/B_{min}$ ~3:1 (ameliorating the dipole design issues of persistent current multipoles and snapback). Separating the injector from the hadron collider ring eliminates problematic issues for the detectors at the intersection regions.

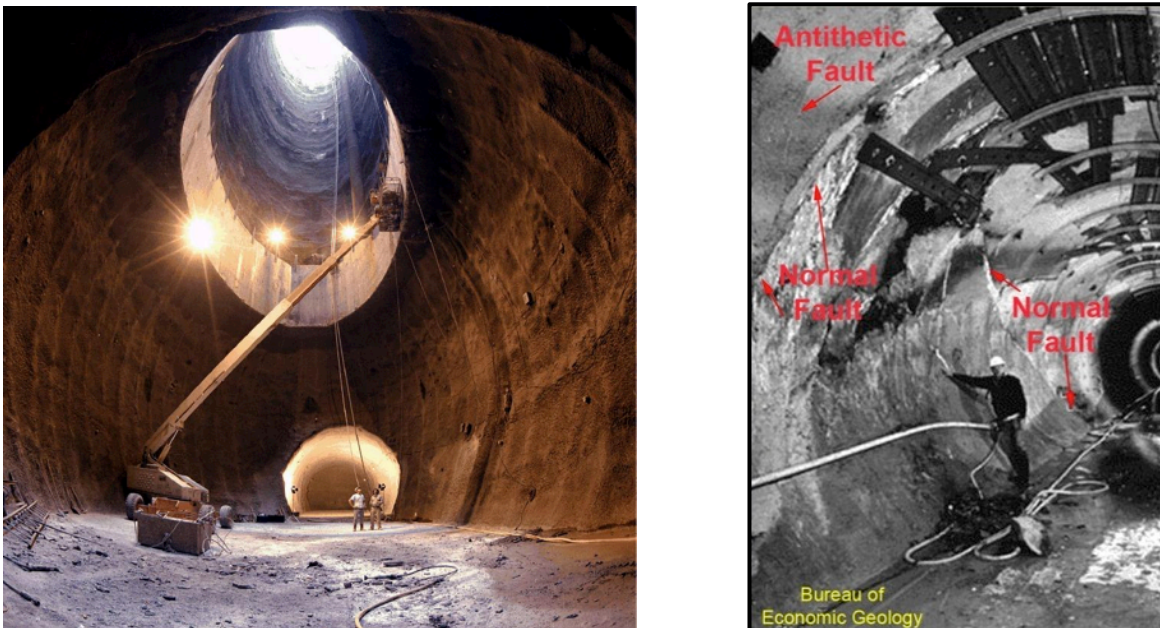

**Figure 1. left) the SSC tunnel and one experimental hall in the Austin chalk and Taylor Marl; right) The unlined tunnel at the location where the tunnel passed through a (passive) fault. Even at that location the world-record advance rate was sustained.**

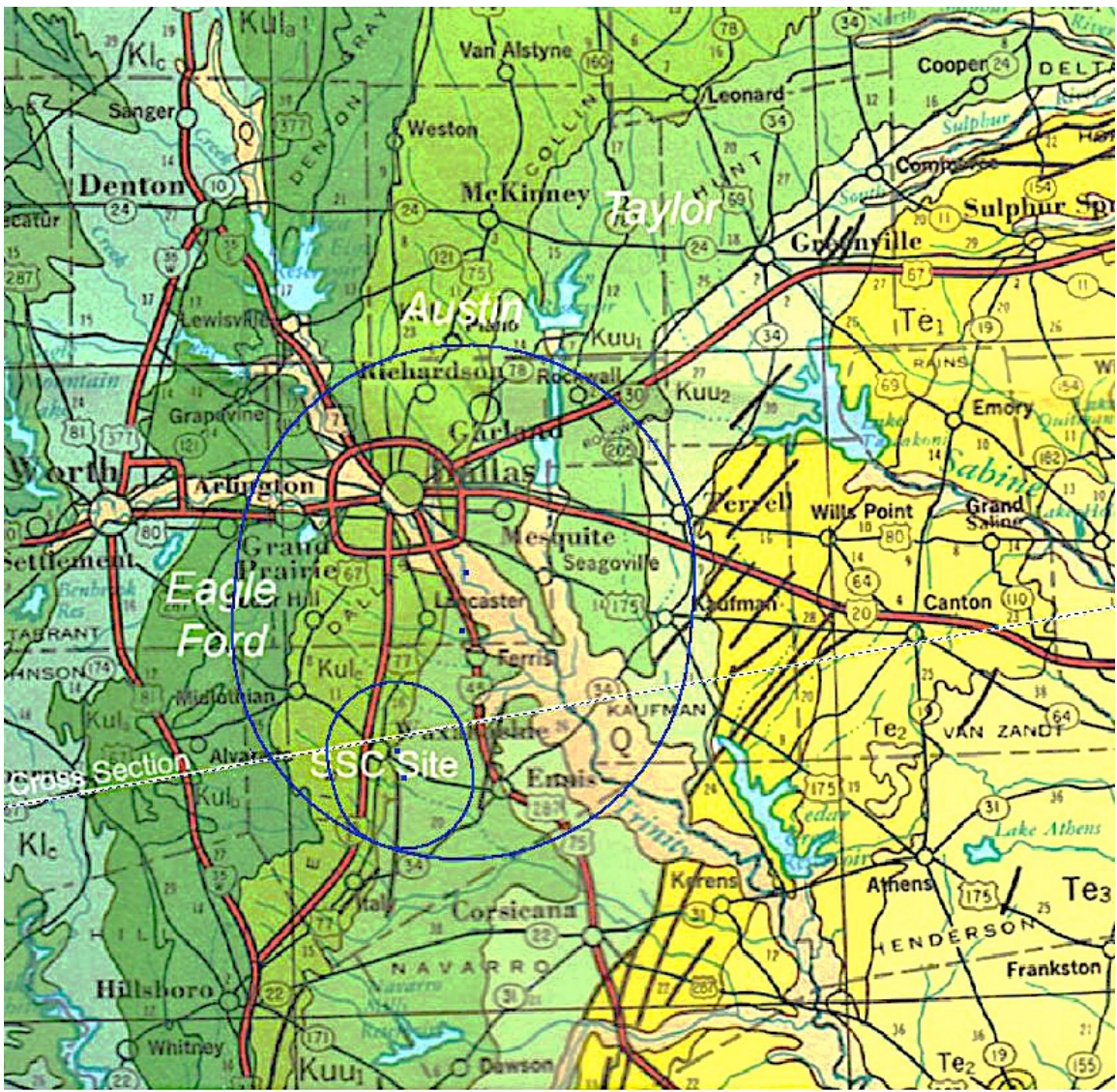

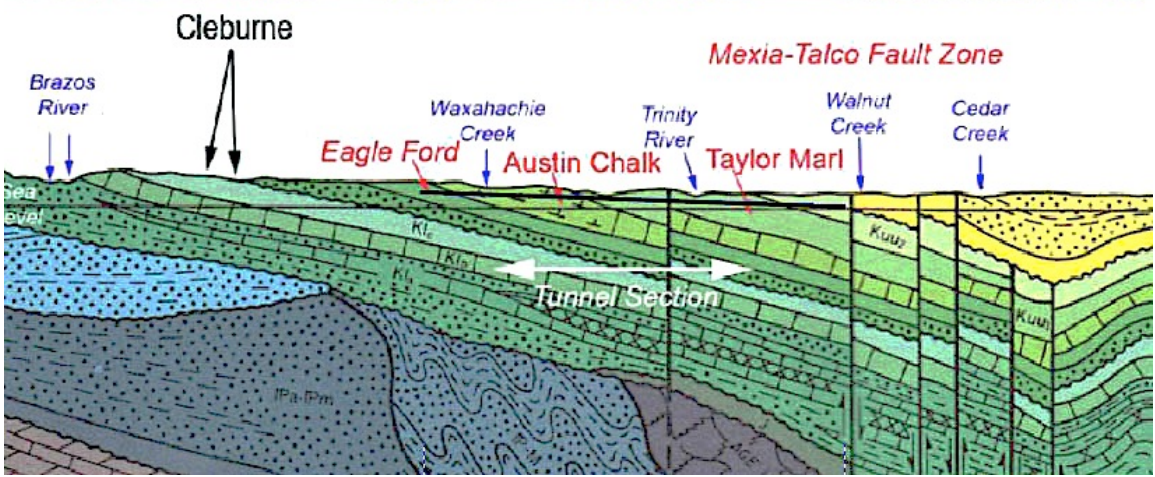

**Figure 2 Geologic map showing the SSC tunnel and an example 270 km circumference tunnel for a 100-300 TeV hadron collider. Both tunnels lie entirely in favorable rock formations for low-cost, rapid-advance-rate tunnel boring.**

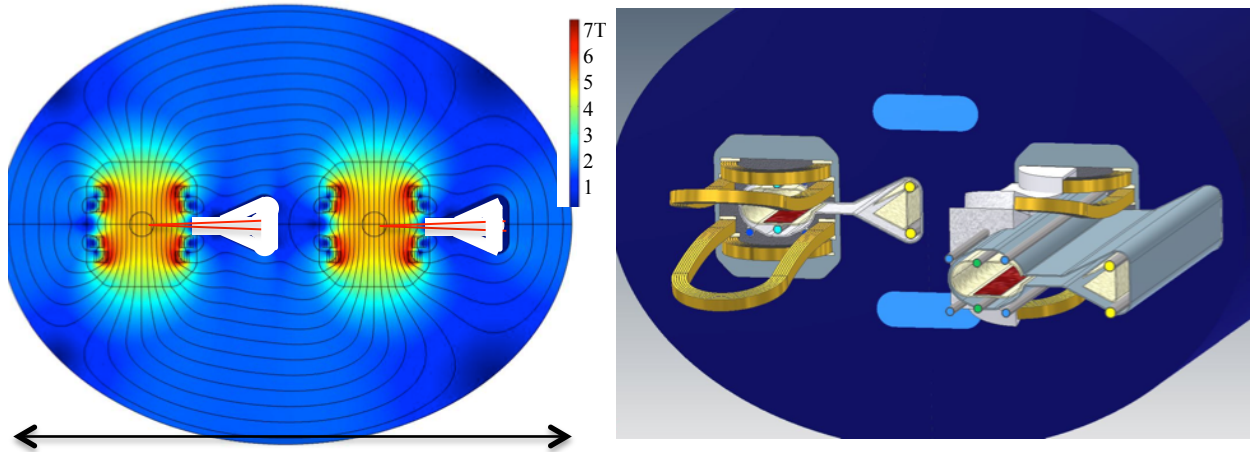

**Figure 3. 5 Tesla C-geometry dual dipole using Nb₃Sn superconducting windings operating at 8 K: a) cross-section showing the calculated field distribution and the side channels for absorbing synchrotron radiation in a separate chamber; b) isometric view of the end winding showing the race-track pancake windings and the flare of the central winding to accommodate the beam tube.**

A 100 TeV hadron collider in the 270 km circumference tunnel would require a dipole field of 4.5 Tesla. An excellent example of a similar dipole is the single-shell cos θ dipole used for RHIC [10], which uses NbTi superconducting cable, operates at 4.5K and produces 3.5 T. It is a nice example of a production-engineered structure, and is likely a global cost-minimum in cost/GeV for a collider dipole. But if one were to press the design to higher field it would require adding a second shell, similar to the dipoles of LHC, and increase the production complexity and cost.

Figure 3 shows our preliminary design for a 5 Tesla C-geometry dipole that can built using simple pancake windings and provides a separate side channel to remove the heat and gas-loading challenges of the intense synchrotron light fan from the beam tube. The dipole can be built with the same cross-section using either NbTi or Nb₃Sn windings, as shown by the operating points in Figure 4. When the dipole is operated at 5 T bore field the maximum field in conductor is ~7.2 T. NbTi has about half the current density of Nb₃Sn under those conditions (Table 1), so the windings would require twice as many turns; otherwise the field designs are identical.

The coil and flux return are cooled by 5K or 8K He flow channels (aqua). Each bore contains a beam screen that is cooled at <60 K by return He vapor flow (green). A clearing electrode is located along the bottom of the beam screen to clear electron cloud. Each dipole is configured as a C geometry so that the horizontal fan of synchrotron radiation escapes the beam tube through a slot aperture and is absorbed in a separate channel. That channel contains NEG vacuum pumping and is cooled at ~150 K by separate cooling channels (yellow). The heat from synchrotron light and the associated gas desorption are removed from the beam tube, and the refrigeration efficiency is improved by a factor >10. The heat and vacuum loads from synchrotron radiation are therefore not an ultimate limit to the achievable luminosity. For the field strength of 5 T, the C geometry shown requires no more superconductor than a conventional dipole. For field choices of 15-20 Tesla required for a 80-100 km tunnel circumference, a C dipole would not be feasible.

All windings are arranged in 4 rectangular block coils, which can be wound as rectangular racetrack windings using automated tooling. The top winding layers remain planar at the ends; the center windings are flared at the ends after windings to accommodate the beam tube. Figure 6 shows the ease of flaring on an actual model winding. The block coil geometry is readily adapted to automated tooling, and was used successfully in the 3 T SSC superferric magnets [11]. The design is being further evolved in collaboration with several industrial firms to optimize it for rapid production and minimum touch labor.

Figure 5 compares the total cross-sectional area of the superconducting wire required for various collider dipole designs, extracting the amount of superconducting wire required in each case from Figure 4. The present-day cost of superconducting wire in large volume is ~$1/kA-m for NbTi, ~$2.5/kA-m for $Nb_3Sn$. Table 2 summarizes the cost of the tunnel (based upon the SSC tunnel costs) and of the superconducting wire for the dipole design presented above, for three cases: the above dipole design for a 270 km tunnel using the options of NbTi or $Nb_3Sn$ superconductor, and a 100 km circumference using 15 T $Nb_3Sn$ dipoles.

The comparison of these two cost drivers makes the impact of the choices clear. The sum of technology and tunnel costs is several times less expensive if one uses NbTi dipoles in a 270 km tunnel. One would choose $Nb_3Sn$ only if the savings in refrigeration power reduced the life-cycle operating cost by more than the discounted premium in capital cost.

Table 3 shows the main parameters for a 100 TeV hadron collider in two cases: for the 100 km circumference tunnel using 15 T dipoles proposed by FCC, and for a 270 km circumference tunnel using 4.5 T dipoles. The parameters highlighted in red are ones that pose particular challenges. *Note that there are no red entries for the 270 km 100 TeV hadron collider.* The salient differences relate to the superconducting magnet technology that is required and the synchrotron radiation power that must be managed within the bore of the superconducting magnets. Recent developments in high-field magnet technology give hope that it should be possible to develop collider dipoles to 15 Tesla and beyond [12], but that same experience has made it clear that they will be difficult and expensive. We propose instead the large-circuference low-field 100 TeV collider, for which proton beams are injected from a 15 TeV ring located separately in the SSC tunnel along with the Higgs Factory. This approach has the nice feature of limiting the dynamic range $B_{max}:B_{min}$ = 3:1 for the large collider, making issues such as persistent-current multipoles more manageable than they would be with a larger dynamic range.

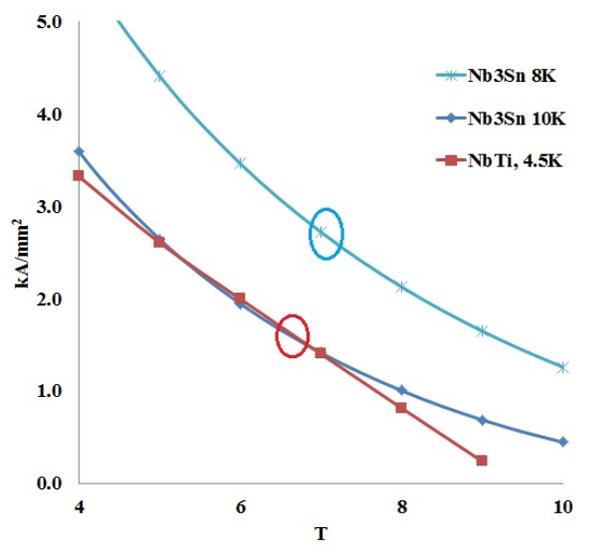 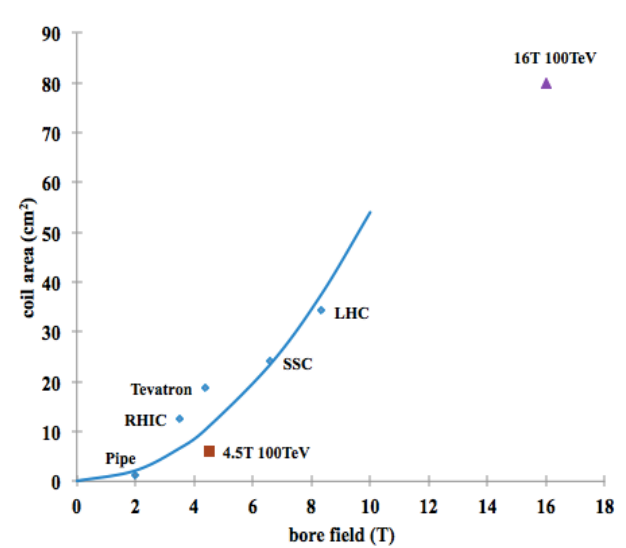

**Figure 4.** Superconductor current density vs. field for NbTi and $Nb_3Sn$. Two candidate choices are NbTi @4.5 K and $Nb_3Sn$ @ 8 K.

**Figure 5.** Cross-section area of superconducting windings for collider dipoles, including a 16 T $Nb_3Sn$ dipole and the present 5 T dipole for a 100 TeV hadron collider (in green).

Table 1. Main parameters of 5 T dual dipole for the Nb$_3$Sn and NbTi coil options.

|  | Nb$_3$Sn option | NbTi option |  |
|---|---|---|---|
| # turns | 6+6 | 11+11 |  |
| Strand diameter | 1.0 | 1.0 | mm |
| Cu/NonCu | 1 | 1 |  |
| # strands in cable | 16 | 16 |  |
| J$_c$ | 2625@ 7.2T,8K | 1500@ 6.5T,4.5K | A/mm$^2$ |
| Max Field in SC | 7.2 | 6.5 | T |
| Bore Field | 5 | 5 | T |
| Stored Energy/bore | 69 | 78 | kJ/m |
| Coil current | 16.5 | 9.4 | kA |
| Self-inductance | 0.51 | 1.7 | mH/m |
| SC strand area/bore | 6.03 | 11.05 | cm$^2$ |

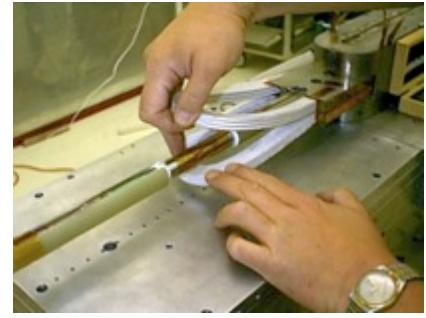

Figure 6. Spreading the inner windings to clear the beam tube.

Table 2. Cost of tunnel and superconductor for dipoles for three 100 TeV hadron collider options.

| Tunnel circumference | 270 km | 270 km | 100 km |
|---|---|---|---|
| Tunnel cost @ $4,900/m | $1320 million | $1320 million | $490 million |
| superconductor material | NbTi @ $1/kA-m | Nb$_3$Sn @ $2.50/kA-m | Nb$_3$Sn |
| GA-m quantity of SC | 740 | 740 | 2800 |
| Superconductor cost | $740 million | $1900 million | $7000 million |

Following Keil's analysis [13] of the dynamics of a synchrotron-radiation-dominated hadron collider, the luminosity $L$ and the total synchrotron radiation power $P$ can be related directly to the total number of protons $N$, the $\beta_x$ at the collision point, and the allowable tune shift $\xi$:

$$L = \frac{\gamma \xi N f}{\beta_x r_p} \qquad P = \frac{8\pi}{3} \frac{N f r e^2}{R} \gamma^4 \qquad L = \left( \frac{3\xi}{8\pi \gamma^3 \beta_x r_p e^2} \right) PR$$

So the achievable luminosity for a given total synchrotron radiation power increases in proportion to collider circumference, and the synchrotron heat per meter deposited in the superconducting magnets for a given luminosity decreases inversely with $R^2$.

The ultimate benefit of a large-circuference option for the 100 TeV hadron collider is that it provides the basis for a future upgrade, if/when 15 T superconducting magnets are demonstrated and affordable, to produce hadron colliding beams with 300 TeV collision energy. The main parameters for the 300 TeV hadron collider are presented in the last column of Table 3. It utilizes the 100 TeV, 270 km ring as injector in the same tunnel, again with dynamic range of 3:1.

The performance of the 300 TeV hadron collider is dominated by effects from synchrotron radiation. Limiting the number of particles per beam to limit synchrotron radiation power produces a short beam lifetime ~3 hours. But synchrotron damping of the vertical emittance makes it possible to do luminosity-leveling by adjusting $\beta_y^*$ to sustain roughly constant tune shift $\xi = .01$. It should be possible to sustain a luminosity of $\sim 10^{35}$ cm$^{-2}$s$^{-1}$ over a ~4 hour period.

Table 3. Main parameters of hadron colliders of 100 and 270 km circumference.

|  | Higgs factory | hadron collider | | | |
|---|---|---|---|---|---|
| Circumference | 100 | 100 | 270 | | km |
| Collision energy | 0.24 | 100 | 100 | 300 | TeV |
| Dipole field | 0.046 | 15 | 4.5 | 14.5 | Tesla |
| Luminosity/I.P. | 5 | 5 | 5 | 10 | $10^{34}$ cm$^{-2}$s$^{-1}$ |
| β* | 50x0.1 | 110 | 50 | 100, 10 | cm |
| Total synch. power | 100 | 4.2 | 1.0 | 34 | MW |
| Critical energy | 430 | 4.0 | 1.0 | 28 | keV |
| Synch power/meter/bore | 580 | 26 | 2 | 80 | W/m |
| Emittance damping time |  | 1 | 19 | .66 | Hr |
| Luminosity lifetime | 0.3 | 18 | 20 | 3.7 | hr |
| Energy loss/turn | 2100 | 4.3 | 1.3 | 114 | MeV |
| RF accel. voltage: | 6000 | 100 | 50 | 250 | MV |
| Acceleration time | .01 |  | .42 | .25 | H |
| Bunch spacing | 250 | 50 | 25 | 25 | ns |
| Beam-beam tune shift | 0.09 | .01 | .01 | .01 |  |
| # IPs | 4 | 2+2 | 2+2 | 2+2 |  |
| # particles per beam | 4.1 | 100 | 220 | 86 | $10^{13}$ |
| Injection energy | 0.12 | >3 | 15 | 50 | TeV |
| Superconducting temp. | 1.8 K in SRF | 4.5 | 8 | 4.5 | K |

**Dual use of the Higgs Factory injector for an X-ray FEL for Structural Systems Biology**

The 9 GeV recirculating linac is required to produce positrons and electrons and to provide initial acceleration for their injection into the Higgs Factory. It also has optimum properties to be used additionally to drive an X-ray FEL for the needs of protein spectroscopy [2]. Consultation with leaders in that area give the following parameters that are needed for their research: 3 mJ pulse energy; 1 kHz pulse rate; 4-12 keV X-ray energy. Those parameters are not available at any present light source, but it should be possible to deliver them using the 9 GeV recirculating linac. The injector can be used for both purposes without sacrificing its performance for either – an ideal dual-use benefit.

**Summary : the minimum-cost path to a Higgs Factory and a 100 TeV Hadron Collider would use the SSC site as a starting point and return HEP leadership to the US.**

The large expanse of favorable rock strata at the SSC site and the existing tunnels already there offer an opportunity to build the facilities proposed in the FCC initiative at minimum cost. The approach described could be implemented using state-of-art technology for superconducting magnets and SRF, with no major risks of technology or cost. The State of Texas once before committed $1 billion as cost-sharing to a high-energy project, and it would be reasonable to hope that it might commit to complete all conventional facilities for the project as cost-sharing. The simple magnet technology offers a basis by which to share the fabrication of the magnets for the hadron collider could be realistically shared among all the industrialized countries of the world, so that the project could be undertaken as a truly World Laboratory.

We have shown that the Higgs Factory and the 100 TeV Hadron Collider are minimum-cost and minimum-risk in the proposed approach, whether they are undertaken together or independently. Siting the next discovery facility in the US would re-energize our field. This is the only such option that has a chance to happen in our professional lifetime.

We have submitted this document to the P5 subpanel of HEPAP, and requested their endorsement of the importance of developing a serious assessment of the proposed facility as a priority for US high energy physics. We plan to organize a special session of the July FCC workshop at Fermilab to develop a collaboration for that purpose.